# Observation of inverse Anderson transitions in Aharonov-Bohm topolectrical circuits

Haiteng Wang[1*], Weixuan Zhang[1*], Houjun Sun[2], and Xiangdong Zhang[1$]

[1] Key Laboratory of advanced optoelectronic quantum architecture and measurements of Ministry of Education, Beijing Key Laboratory of Nanophotonics & Ultrafine Optoelectronic Systems, School of Physics, Beijing Institute of Technology, 100081, Beijing, China

[2] Beijing Key Laboratory of Millimeter wave and Terahertz Techniques, School of Information and Electronics, Beijing Institute of Technology, Beijing 100081, China

*These authors contributed equally to this work.

$Author to whom any correspondence should be addressed. E-mail: zhangxd@bit.edu.cn

**It is well known that Anderson transition is a disorder-induced metal-insulator transition. Contrary to this conventional wisdom, some investigations have shown that disorders could destroy the phase coherence of localized modes in flatbands, making the localized states melt into extended states. This phenomenon is called the inverse Anderson transition. While, to date, the experimental observation of inverse Anderson transitions is still lacking. In this work, we report the implementation of inverse Anderson transitions based on Aharonov-Bohm topolectrical circuits. Different types of disorders, including symmetric-correlated, antisymmetric-correlated and uncorrelated disorders, can be easily implemented in Aharonov-Bohm circuits by engineering the spatial distribution of ground settings. Through the direct measurements of frequency-dependent impedance responses and time-domain voltage dynamics, the inverse Anderson transitions induced by antisymmetric-correlated disorders are clearly observed. Moreover, the flat bands and associated spatial localizations are also fulfilled in clean Aharonov-Bohm circuits or Aharonov-Bohm circuits sustaining symmetric-correlated and uncorrelated disorders, respectively. Our proposal provides a flexible platform to investigate the interplay between the geometric localization and Anderson localization, and could have potential applications in electronic signal control.**

Anderson localization describes the complete localization of non-interacting particles induced by the uncorrelated disorder that is added into a regular lattice [1]. Such a localization is the direct consequence of interference between different paths arising from multiple scattering of the electron by lattice defects. Owing to the exotic phenomenon of Anderson localization, there are many investigations on the interplay between Anderson localizations and other novel physics, such as the disorder-induced topological insulators [2-7] and mobility edges [8-10]. Contrary to the Anderson localization, some investigations also show that disorders can induce localization-delocalization transition in the system with all bands being flat bands [11]. Such a novel phenomenon is called the inverse Anderson localization and was originally predicted in the three-dimensional diamond lattice with four-fold degenerated orbitals [11]. While, up to now, the experimental observation of inverse Anderson localizations is still not reported in any classical or quantum systems. How to detect such a novel phenomenon is still an open question.

Except for the disorder-induced Anderson localization, Aharonov-Bohm cages correspond to another localization effect, which is induced by the destructive interferences of electron beams traveling along paths enclosing a magnetic flux. The first theoretical work discovered Aharonov-Bohm cages were based on the lattice model sustaining Dice tiling and Hofstadter butterfly at a critical value for the magnetic flux per plaquette [12]. Lately, the exotic Aharonov-Bohm cages are subsequently extended in diamond chain [13, 14] and experimentally verified in different quantum and classical systems [15-19]. Following the experimental realization of Aharonov-Bohm cages, a variety of related theoretical works has been presented to further reveal many novel effects related to Aharonov-Bohm cages, including the influence of nonlinearity and disorder on the Aharonov-Bohm localization [20, 21], topological pumping and edge states [22], the non-Hermitian Aharonov-Bohm cages [23] and so on [24-27]. Very recently, Longhi theoretically shows that the antisymmetric-correlated disorder could induce inverse Anderson localizations in Aharonov-Bohm cages, where disorders can in principle induce a localization-delocalization transition [27]. This discovery further enriches the available platform to observe inverse Anderson localizations in experiments.

In this work, we give the first experimental investigation of inverse Anderson transitions based on Aharonov-Bohm topolectrical circuits. By mapping the eigenmodes of one-dimensional Aharonov-Bohm cage to the modes of the designed circuit simulators, the strong voltage localization is verified by designed Aharonov-Bohm circuit. Based on the flexibly of the circuit connections and groundings, different types of disorders, including the symmetric-correlated, antisymmetric-correlated and uncorrelated disorders, can be easily implemented in the Aharonov-Bohm circuits. Thus, the disordered Aharonov-Bohm circuit provides an ideal platform to observe novel effects resulting from the interplay between geometric localizations and disorders. It is found that flat bands and the associated localization effect are still maintained in Aharonov-Bohm circuits sustaining symmetric-correlated and uncorrelated disorders, respectively. In addition, the antisymmetric-correlated disorder induced inverse Anderson transitions are clearly observed through the direct measurements of frequency-dependent impedance responses and time-domain voltage dynamics.

**The theory of inverse Anderson transitions in Aharonov-Bohm topolectrical circuits.** We consider a one-dimensional lattice model, where three sublattices (as marked by '$a$', '$b$' and '$c$') exist in a single unit cell, and a synthetic magnetic flux is applied in each closed loop via Peierls' substitution of the coupling strength ($Je^{i\theta}$) between sublattices $a$ and $b$, as shown in Fig. 1a. In this case, the system can be described by the Hamiltonian as:

$$H = \sum_{n=1}^{N} J(e^{i\theta} a_n^+ b_n + a_n^+ c_n + a_{n+1}^+ b_n + a_{n+1}^+ c_n + h.c.) + U_a a_n^+ a_n + U_b b_n^+ b_n + U_c c_n^+ c_n \quad (1)$$

where the operators $a_n^+$, $b_n^+$ and $c_n^+$ ($a_n$, $b_n$, $c_n$) correspond to the creation (annihilation) operators of three different sublattices in the unit cell labeled by $n$. $N$ is the total number of unit cells in the periodic 1D lattice. $J$ is the hopping rate between adjacent sites. The phase factor $\theta$ equals to the synthetic magnetic flux in a single unit. $U_a$, $U_b$, $U_c$ defines the on-site energy of three lattice sites $a$, $b$ and $c$, respectively. The probability amplitude of a single particle hopping on the 1D lattice can be expressed as $|\psi\rangle = \sum_{n=1}^{N}(\varphi_{a_n} a_n^+ + \varphi_{b_n} b_n^+ + \varphi_{c_n} c_n^+)|0\rangle$ where $|0\rangle$ is the vacuum state and $\varphi_{a_n}$, $\varphi_{b_n}$ and $\varphi_{c_n}$ are the probability amplitude with a single particle at the site $a$, $b$, $c$ of the $n$th cell. Solving the Schrödinger equation $H|\psi\rangle = \varepsilon|\psi\rangle$, we obtain the eigen-equation with respect to $\varphi_{a_n}$, $\varphi_{b_n}$ and $\varphi_{c_n}$ as:

$$\varepsilon \varphi_{a_n} = J(e^{i\theta}\varphi_{b_n} + \varphi_{b_{n-1}} + \varphi_{c_n} + \varphi_{c_{n-1}}) + U_a \varphi_{a_n} \quad (2)$$

$$\varepsilon \varphi_{b_n} = J(e^{-i\theta}\varphi_{a_n} + \varphi_{a_{n+1}}) + U_b \varphi_{b_n} \quad (3)$$

$$\varepsilon \varphi_{c_n} = J(\varphi_{a_n} + \varphi_{a_{n+1}}) + U_c \varphi_{c_n} \quad (4)$$

In this case, the lattice sustains three bands with dispersion relations given by $\varepsilon_0 = 0, \varepsilon_\pm =$

$\pm 2J\sqrt{1 + \cos(\theta/2)\cos(q + \theta/2)}$ where $-\pi \leq q < \pi$ is the Bloch wavevector. For $\theta = \pi$, there will be three flat bands with the corresponding eigen-energies being $\varepsilon_0 = 0$ and $\varepsilon_\pm = \pm 2J$. The localized eigenstates associated to these three flat bands are displayed in Figure 1(b), where the mode distribution is the same with different values of Bloch wavevector $q$. Due to the flat-band dispersion and the strong localization of eigenmodes, the propagation of any wave packet is suppressed in the lattice, corresponding to the so-called Aharonov-Bohm cage. There are many theoretical investigations on the influence of different types of disorders on the localization effects of Aharonov-Bohm cage. While, to date, a suitable experimental platform to fulfill these novel effects is still lacking.

Based on the similarity between circuit Laplacian and lattice Hamiltonian [28-47], electric circuits can be used as an ideal platform to achieve the Aharonov-Bohm cage with disorders. Here, a pair of circuit nodes connected by the capacitor $C$ are considered to form an effective site in the lattice model. Voltages at these two nodes in the *n*th unit are expressed as $V_{\alpha_n,1}$ and $V_{\alpha_n,2}$ ($\alpha_n = a_n, b_n, c_n$), which are suitably formulated to form a pair of pseudospins $V_{\uparrow\alpha_n,\downarrow\alpha_n} = V_{\alpha_n,1} \pm V_{\alpha_n,2}$. To simulate the real-valued hopping rate, two capacitors $C$ are used to directly link adjacent nodes without a cross, as shown in Figure 1(c). As for the realization of hopping rate with a phase ($e^{\pm i\pi}$), two pairs of adjacent nodes are cross connected via $C$, as plotted in Figure 1(d). In addition, each node is grounded by an inductor $L_g$. To ensure the same resonance frequency, the extra capacitor $2C$ should be used to connect subnodes *b* and *c* to the ground. Moreover, three types of capacitors $C_a$, $C_b$, $C_c$ are used for grounding on three subnodes to simulate the onsite energy. Finally, a capacitor $C_u$ is grounded on each node, corresponding to an effective translation of eigen-spectrum of the system. Through the appropriate setting of grounding and connecting, the circuit eigen-equation can be derived as:

$$\left(\frac{f_0^2}{f^2} - 6 - \frac{C_u}{C}\right)V_{\downarrow,a_n} = -\left(e^{i\pi}V_{\downarrow,b_n} + V_{\downarrow,b_{n-1}} + V_{\downarrow,c_n} + V_{\downarrow,c_{n-1}}\right) + \frac{C_a}{C}V_{\downarrow,a_n} \quad (5)$$

$$\left(\frac{f_0^2}{f^2} - 6 - \frac{C_u}{C}\right)V_{\downarrow,b_n} = -\left(e^{-i\pi}V_{\downarrow,a_n} + V_{\downarrow,a_{n+1}}\right) + \frac{C_b}{C}V_{\downarrow,b_n} \quad (6)$$

$$\left(\frac{f_0^2}{f^2} - 6 - \frac{C_u}{C}\right)V_{\downarrow,c_n} = -\left(V_{\downarrow,a_n} + V_{\downarrow,a_{n+1}}\right) + \frac{C_c}{C}V_{\downarrow,c_n} \quad (7)$$

where $f$ is the eigen-frequency ($f_0 = 1/2\pi\sqrt{L_g C}$) of the designed circuit, and $V_{\downarrow,\alpha_n} = (V_{\alpha_n,1} - V_{\alpha_n,2})/\sqrt{2}$ represents the voltage pseudospin. Details for the derivation of circuit eigen-equations are provided in [48]. It is shown that the eigen-equation of the designed electric circuit possesses the same form with Eqs. (2)-(4). In particular, the probability amplitude for the 1D lattice model $\varphi_{a_n}$, $\varphi_{b_n}$ and $\varphi_{c_n}$ are mapped to the voltage of pseudospin $V_{\downarrow,a_n}$, $V_{\downarrow,b_n}$ and $V_{\downarrow,c_n}$. The eigen-energy ($\varepsilon$) of Aharonov-Bohm cage is directly related to the eigen-frequency ($f$) of the circuit as $\varepsilon = f_0^2/f^2 - 6 - C_u/C$. And, other parameters are given by $J = -1$ and $U_{a,b,c} = C_{a,b,c}/C$. In this case, we have achieved the above proposed Aharonov-Bohm cage in electric circuits.

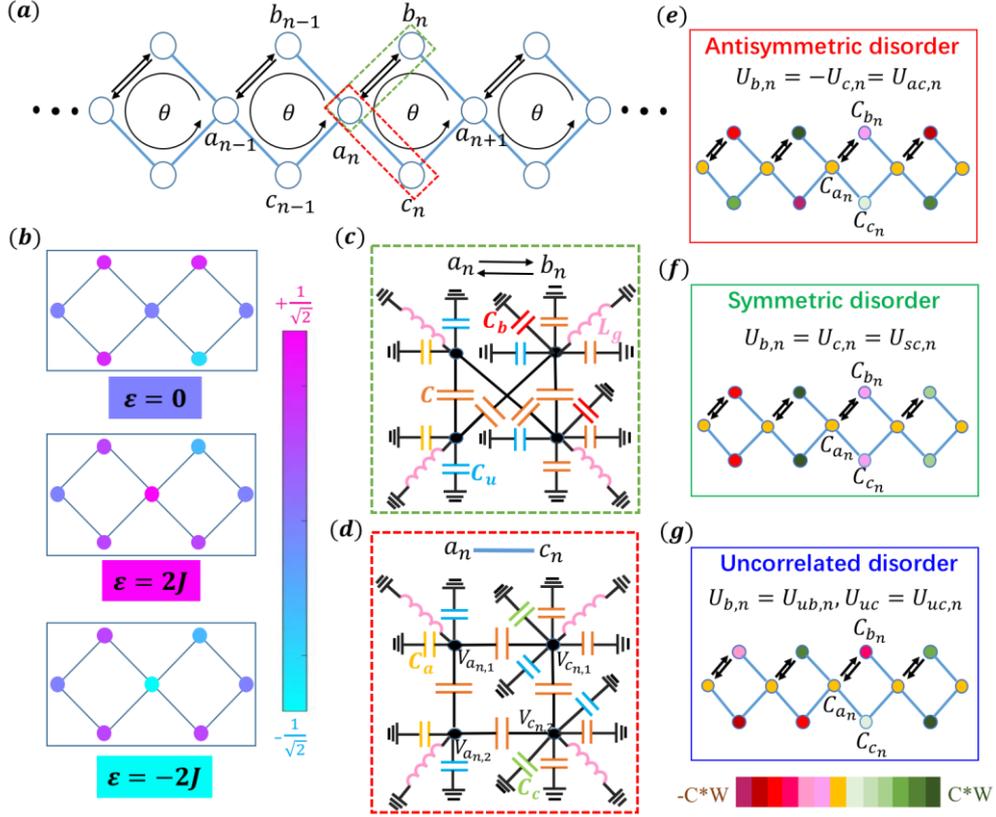

**FIG. 1.** (a) Schematic diagram of the quasi 1D lattice with periodic boundary conditions. The unit-cell contains three sublattices marked by *a*, *b* and *c*. A synthetic magnetic flux with $\theta = \pi$ is applied in each closed loop. (b) Spatial distributions of three localized eigenstates with eigen-energies being $\varepsilon_0 = 0$ and $\varepsilon_\pm = \pm 2J$. (c) and (d). Illustrations of complex-valued and real-valued hopping rates in circuits between two sites enclosed by green and red dash blocks in Fig. 1(a). (e)-(g) The sketch map for the antisymmetric-correlated disorder, symmetric-correlated disorder, and uncorrelated disorder of the designed electric circuit. The grounding capacitors $C_{b_n}$ and $C_{c_n}$ at different units take required random values of $C*[-W, W]$ to fulfill the required disordered potential.

To investigate the interplay between disorder effects and Aharonov-Bohm cage, it is important to introduce different types of disorders into the designed electric circuit. Actually, we consider three kinds of disorders added on sublattices *b* and *c*. The first is called the symmetric-correlated disorder, which corresponds to the case with onsite energies of sublattices *b* and *c* in different unit cells being $U_{b,n} = U_{c,n} = U_{sc,n}$. $U_{sc,n}$ is the independent random number in the range of [-W, W].

The second is called the antisymmetric-correlated disorder, where the onsite energies of sublattices b and c in different unit cells is in the form of $U_{b,n} = -U_{c,n} = U_{ac,n}$ with $U_{ac,n}$ being random numbers in the range of [-W, W]. The last one is the uncorrelated disorder, which can be realized by controlling the onsite energies of sublattices b ($U_{b,n} = U_{ub,n}$) and c ($U_{c,n} = U_{uc,n}$) in different units. Here b and c are two independent random numbers in the range of [-W, W]. It is worthy to note that these three types of disorders could be easily introduced into the designed electric circuit by setting the grounding capacitors $C_{b_n}$ and $C_{c_n}$ at different units as required random values $C*$[-W, W]. In this case, the schematic diagrams of electric circuits with antisymmetric-correlated, symmetric-correlated and uncorrelated disorders are shown in Figs. 1(e), 1(f), and 1(g). In this case, to avoid the appearance of negative capacitors in the grounding of disordered circuits, the value of capacitor $C_u$ should be larger than $C*W$.

To analyze the interplay between three types of disorders and Aharonov-Bohm cage, we firstly calculate the eigen-spectra and the inverse participation ratio (IPR) of each eigenmode in designed electric circuit with different disorders, as shown in Figure 2. Here, the IPR is defined as $IPR = \sum_{n=1}^{N} \left( |V_{\downarrow,a_n}|^4 + |V_{\downarrow,b_n}|^4 + |V_{\downarrow,c_n}|^4 \right)$. We note that the IPR of an extended state scales as $1/N$, and vanishes in the large $N$ limit. While, it remains finite for a localized state. In the following calculation, circuit parameters are set as $C = 1nf$, $L_g = 10uH$, $C_u = 1nf$, and $W = 0.25$, respectively. Without loss of generality, we delete the grounding capacitor $C_{a_n}$ in the designed circuit. As shown in Fig. 2(a), the eigen-frequencies of electric circuit without disorders equal to three different values (*f*=0.712*MHz*, 0.602*MHz*, 0.531*MHz*), exhibiting the flat-band effect. These three eigen-frequencies are matched to eigen-energies ($\varepsilon = -2, 0, 2$) of the lattice model of Aharonov-Bohm cage. In addition, IPRs show relatively large values, indicating the strong localization of associated eigenmodes. By introducing antisymmetric-correlated disorders, as shown in Fig. 2(b), it is found that the eigenspectrum is broadening with the appearance of disorder-induced eigen-modes in gap regions of the clean Aharonov-Bohm cage. And the associated IPRs of these eigenmodes are extremely small, corresponding to the extended states. Such a disorder-induced localization-extension transition is called inverse Anderson transition, which is contrary to the effect of Anderson localization with disorder-assisted localizations. Figs. 2(c) and 2(d) show the behaviors of eigen-spectra and IPRs for electric circuits with symmetric-correlated disorder and uncorrelated disorder, respectively. In these cases, we can see the IPRs still keep large values, and there is no significant extension of eigen-spectra under these two types of disorders. We call this phenomenon as disorder-immune localization in Aharonov-Bohm cage. From above results, we can see that only the antisymmetric-correlated disorder can induce a localization-extension transition, while the localization of Aharonov-Bohm cage is robust against the symmetric-correlated and uncorrelated disorders. Physically, only the antisymmetric-correlated disorders on subnodes $b$ and $c$ could significantly break the destructive interference in the Aharonov-Bohm cage, which results in the extension of eigenmodes and induces inverse Anderson transition.

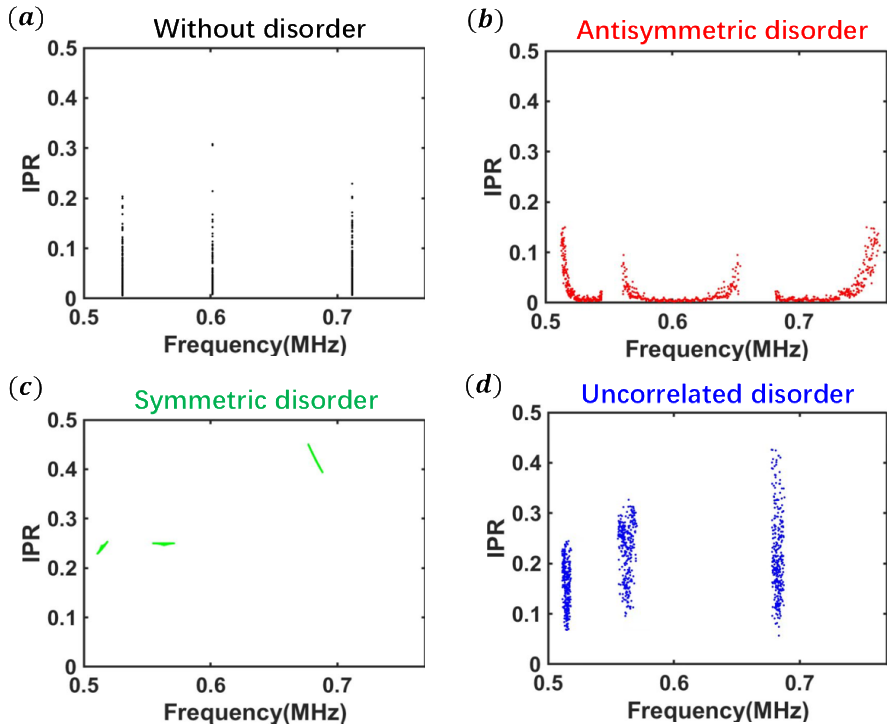

**FIG. 2.** The calculated eigen-spectra and inverse participation ratios of Aharonov-Bohm electric circuits without disorders for (a), and with antisymmetric-correlated, symmetric-correlated and uncorrelated disorders for (b)-(d). Circuit parameters are set as $C = 1nf$, $L_g = 10uH$, $C_u = 1nf$, and $W = 0.25$.

**Experimental observation of inverse Anderson transitions in Aharonov-Bohm electric circuit.** To experimentally observe the inverse Anderson transitions, the designed Aharonov-Bohm circuits are fabricated, and the corresponding parameters are the same to that used in Fig. 2. The photograph image of the circuit sample is presented in Fig. 3(a) and the enlarged views are plotted in right insets. Here a single printed circuit board (PCB), which contains 21 node pairs, is applied for the circuit. The capacitors $C$ connecting a pair of circuit nodes (equivalent to a single lattice site) are framed by orange circles. In this case, if two pair of adjacent circuit nodes are directly (cross) connected through the capacitors $C$ framed by orange squares, the hopping rate without (with) a phase-factor $e^{\pm i\pi}$ could be realized. The inductor $L_g$ and capacitor $C_u$ are enclosed by the pink and blue frames. As circled by yellow squares, two extra capacitors $C$ are used to ground on subnodes $b$ and $c$ to ensure the same resonance frequency with the subnodes $a$. The grounding capacitor $C_{b_n}$ and $C_{c_n}$ (framed by red and green squares) are used to simulate different types of disorders.

It is known that the impedance responses of circuits in the frequency domain are related to the local density of states of the corresponding quantum lattice model. In this case, to evaluate the change of eigen-spectra with different types of disorders, we firstly measure the impedance response of circuit nodes *a, b* and *c* of a unit cell. The solid lines in fig. 3(b) present the corresponding experimental results without disorders. We can see that the impedance peaks at different nodes all locate at three frequencies, matched to the eigen-spectrum of lattice model in Fig. 2(a). The corresponding simulation results are plotted in Fig. 3(b) by dashed lines based on the LTSPICE software. A good consistence between simulations and measurements is obtained, and the larger width of measured impedance peaks results from the lossy effect in fabricated circuits. In addition, we can see that there are two impedance peaks of node *a* and three impedance peaks of nodes *b* and *c*. These phenomena are also consistent with the spatial distribution of eigenstates at different energies, where the zero-energy eigenstates are mainly located at subnodes *b* and *c*, and the eigenstates with eigenvalues being $\pm 2$ are distributed in all three subnodes, as shown in Fig. 1(b). By introducing antisymmetric-correlated disorder, as shown in Fig. 3(c) by solid lines for measurements and dashed lines for simulations, we can see that there are many redundant impedance peaks at different frequencies in gap regions between three main peaks. In addition, the maximum peak value is also decreased compared to circuit without disorders. These results clearly illustrate the extension of eigen-spectrum as well as the delocalization of the corresponding eigen-fields. Furthermore, we also measure and simulate the impedance responses of circuit with symmetric-correlated (uncorrelated) disorders, as shown in Figs. 3(d) (Figs. 3(e)) by solid lines and dashed lines. It is shown that the symmetric-correlated disorder can only make a shift of the impedance peaks, and the peak values still keep the same with that of the clean circuit. Similarly, with the introducing of uncorrelated disorders, there are still three impedance peaks with the corresponding frequencies being unchanged compared to the circuit without disorders, and the peak values nearly keep the same. In this case, the frequency-dependent impedance measurements clearly manifest the change of eigen-spectra and eigen-fields under different types of disorders in circuits. Thus, the antisymmetric-correlated disorder induced inverse Anderson transition, and symmetric-correlated/uncorrelated disorders immune locations are clearly observed.

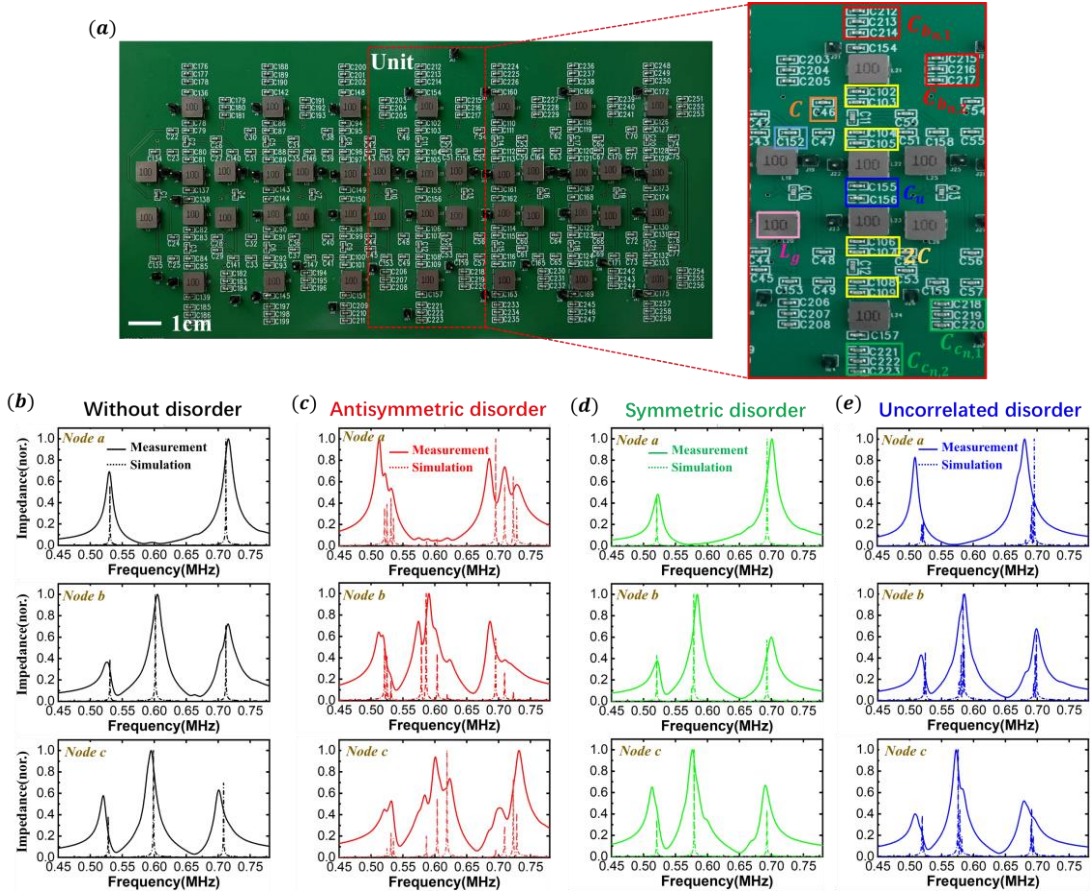

**FIG. 3.** (a) The photograph images of the fabricated circuit simulator. The enlarged views of a unit cell are shown in right insets. (b)-(e) illustrate the impedance responses the circuits without disorders, with antisymmetric-correlated disorders, symmetric-correlated disorder and uncorrelated disorders, respectively, The solid and dash lines correspond to measurements and simulations.

To further experimentally demonstrate the phenomena that only the antisymmetric-correlated disorder can induce the inverse Anderson transitions, in the following, we perform the measurements of temporal dynamics for injected voltage signals with different types of disorders. Here, the circuit excitation is in the form of $[V_{a_4,1} = V_0 e^{i\omega t}, V_{a_4,2} = -V_0 e^{i\omega t}]$, and the associated excitation frequency is set as 0.60MHz, which matches to the eigen-frequency of designed circuits. Details of the experimental measurements are provided in [48].

Fig. 4(a) displays the measured voltage signal of the circuit without disorders in the time-domain. We can clearly see that in the clean circuit, the extremely high voltage signal concentrates at circuit nodes of $b_{3_{1,2}}$, $c_{3_{1,2}}$, $a_{4_{1,2}}$, $b_{4_{1,2}}$ and $c_{4_{1,2}}$, corresponding to the localization in

Aharonov-Bohm cage. Then we measure the time dynamics of voltage signal in the circuit possessing antisymmetric-correlated disorders, as shown in Fig. 4(b). It is clearly shown that the input voltages could exhibit the fast-extension behavior under the influence of antisymmetric-correlated disorders, which destroy the destructive interference in the original Aharonov-Bohm cage and make the voltage signal get spread. Figs. 4(c) and 4(d) present the measured voltage evolutions in the circuit with symmetric-correlated and uncorrelated disorders, respectively. Similar to the clean Aharonov-Bohm circuit, the injected voltage still localized around the input node, indicating the disorder-immune localization. These experimental results are also consistent with simulation results [48].

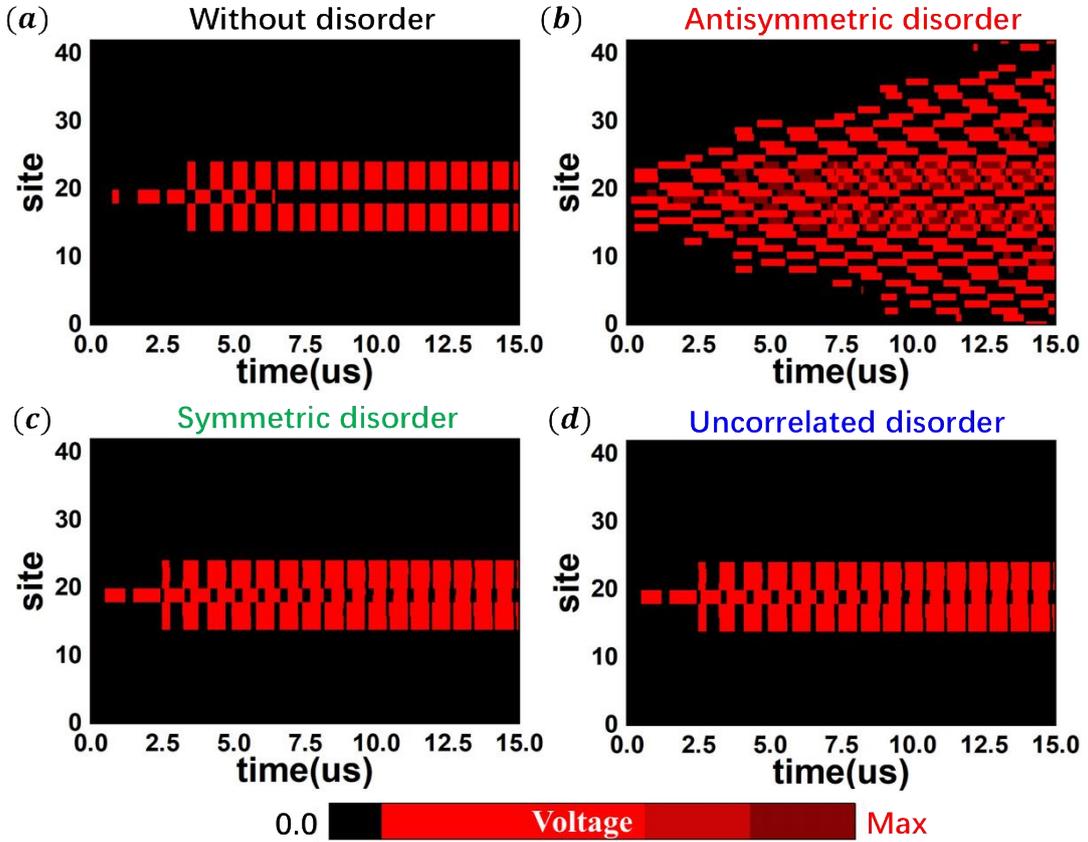

**FIG. 4.** Experimental results of the temporal dynamics for the fabricated Aharonov-Bohm circuits without disorders for (a) and with antisymmetric-correlated, symmetric-correlated disorder and uncorrelated disorders for (b)-(d).

**Conclusion.** In conclusion, we experimentally investigate the inverse Anderson transitions in Aharonov-Bohm circuit networks. The eigenmodes of Aharonov-Bohm cage are mapped to modes of the designed circuit simulators. Based on the Aharonov-Bohm circuit, the strong localization effect is verified. In addition, by engineering the spatial distribution of ground settings, we introduce different types of disorders, including the symmetric-correlated, antisymmetric-correlated and uncorrelated disorders, into the Aharonov-Bohm circuits. Through the direct measurements of frequency-dependent impedance responses and time-domain voltage dynamics, the antisymmetric-correlated disorder induced inverse Anderson transitions are clearly observed. Moreover, the flat bands and the associated localization effects are also fulfilled in the Aharonov-Bohm circuits sustaining symmetric-correlated and uncorrelated disorders, respectively. Our proposal provides a flexible platform to investigate and visualize many interesting phenomena related to disordered Aharonov-Bohm cage and could have potential applications in the field of the electronic signal control.

## References

[1]  P. W. Anderson, Phys. Rev. **109**, 1492 (1958).
[2]  J. Li, R.-L. Chu, J. K. Jain, and S.-Q. Shen, Phys. Rev. Lett. **102**, 136806 (2009).
[3]  H. Jiang, L. Wang, Q.-f. Sun, and X. C. Xie, Phys. Rev. B **80**, 165316 (2009).
[4]  C. W. Groth, M. Wimmer, A. R. Akhmerov, J. Tworzydło, and C. W. J. Beenakker, Phys. Rev. Lett. **103**, 196805 (2009).
[5]  H. M. Guo, G. Rosenberg, G. Refael, and M. Franz, Phys. Rev. Lett. **105**, 216601 (2010).
[6]  A. Altland, D. Bagrets, L. Fritz, A. Kamenev, and H. Schmiedt, Phys. Rev. Lett. **112**, 206602 (2014).
[7]  P. Titum, N. H. Lindner, M. C. Rechtsman, and G. Refael, Phys. Rev. Lett. **114**, 056801 (2015).
[8]  J. T. Chalker, T. S. Pickles, and P. Shukla, Phys. Rev. B **82**, 104209 (2010).
[9]  S. Flach, D. Leykam, J. D. Bodyfelt, P. Matthies, and A. S. Desyatnikov, Europhys. Lett. **105**, 30001 (2014).
[10] J. D. Bodyfelt, D. Leykam, C. Danieli, X. Yu, and S. Flach, Phys. Rev. Lett. **113**, 236403 (2014).
[11] M. Goda, S. Nishino, and H. Matsuda, Phys. Rev. Lett. **96**, 126401 (2006).
[12] J. Vidal, R. Mosseri, and B. Douçot, Phys. Rev. Lett. **81**, 5888 (1998).
[13] M. Creutz, Phys. Rev. Lett. **83**, 2636 (1999).
[14] J. Vidal, B. Douçot, R. Mosseri, and P. Butaud, Phys. Rev. Lett. **85**, 3906 (2000).
[15] S. Mukherjee, M. Di Liberto, P. Öhberg, R. R. Thomson, and N. Goldman, Phys. Rev. Lett. **121**, 075502 (2018).
[16] C. C. Abilio, P. Butaud, T. Fournier, B. Pannetier, J. Vidal, S. Tedesco, and B. Dalzotto, Phys. Rev. Lett. **83**, 5102 (1999).
[17] I. M. Pop, K. Hasselbach, O. Buisson, W. Guichard, B. Pannetier, and I. Protopopov, Phys. Rev. B **78**, 104504 (2008).
[18] G. Möller and N. R. Cooper, Phys. Rev. Lett. **108**, 045306 (2012).
[19] A. Bermudez, T. Schaetz, and D. Porras, Phys. Rev. Lett. **107**, 150501 (2011).
[20] M. Di Liberto, S. Mukherjee, and N. Goldman, Phys. Rev. A **100**, 043829 (2019).
[21] G. Gligorić, D. Leykam, and A. Maluckov, Phys. Rev. A **101**, 023839 (2020).
[22] M. Kremer, I. Petrides, E. Meyer, M. Heinrich, O. Zilberberg, and A. Szameit, Nat. Commun. **11**, 907 (2020).
[23] S. M. Zhang and L. Jin, arXiv:2005.01044 (2020).
[24] J. Vidal, P. Butaud, B. Douçot, and R. Mosseri, Phys. Rev. B **64**, 155306 (2001).
[25] C. E. Creffield and G. Platero, Phys. Rev. Lett. **105**, 086804 (2010).
[26] J. Zurita, C. E. Creffield, and G. Platero, Adv. Quantum Technol. **3**, 1900105 (2020).

[27] S. Longhi, Opt. Lett. **46**, 2872 (2021).

[28] J. Ningyuan, C. Owens, A. Sommer, D. Schuster, and J. Simon, Phys. Rev. X **5**, 021031 (2015).

[29] V. V. Albert, L. I. Glazman, and L. Jiang, Phys. Rev. Lett. **114**, 173902 (2015)..

[30] C. Lee et al., Commun. Phys. 1, 39 (2018).

[31] S. Imhof *et al.*, Nat. Phys. **14**, 925 (2018).

[32] M. Ezawa, Phys. Rev. B **100**, 075423 (2019).

[33] T. Helbig, T. Hofmann, C. H. Lee, R. Thomale, S. Imhof, L. W. Molenkamp, and T. Kiessling, Phys. Rev. B **99**, 161114 (2019).

[34] Y. Lu, N. Jia, L. Su, C. Owens, G. Juzeliūnas, D. I. Schuster, and J. Simon, Phys. Rev. B **99**, 020302 (2019).

[35] W. Zhang, D. Zou, J. Bao, W. He, Q. Pei, H. Sun, and X. Zhang, Phys. Rev. B **102**, 100102 (2020).

[36] L. Song, H. Yang, Y. Cao, and P. Yan, Nano Lett. **20**, 7566 (2020).

[37] Y. Wang, H. M. Price, B. Zhang, and Y. D. Chong, Nat. Commun. **11**, 2356 (2020).

[38] N. A. Olekhno *et al.*, Nat. Commun. **11**, 1436 (2020).

[39] W. Zhang, D. Zou, Q. Pei, W. He, J. Bao, H. Sun, and X. Zhang, Phys. Rev. Lett. **126**, 146802 (2021).

[40] M. D. Ventra, Y. V. Pershin, and C. Chien, Phys. Rev. Lett. 128, 097701 (2022).

[41] D. Zou, T. Chen, W. He, J. Bao, C.H. Lee, H. Sun, and X. Zhang, Nat. Commun. **12** 7201 (2021).

[42] S.S. Yamada *et al., Nat Commun* **13,** 2035 (2022).

[43] A. Stegmaier *et al.*, Phys. Rev. Lett. **126**, 215302 (2021).

[44] W. Zhang, F. Di, H. Yuan, H. Wang, X. Zheng, L. He, H. Sun, and X. Zhang, Phys. Rev. B **105,** 195131 (2022).

[45] S. Liu, R. Shao, S. Ma, L. Zhang, O. You, H. Wu, Y. J. Xiang, T. J. Cui, and S. Zhang, Research **2021**, 5608038 (2021).

[46] W. Zhang, H. Yuan, H. Wang, F. Di, N. Sun, X. Zheng, H. Sun, and X. Zhang, Nat. Commun. **13**, 2392 (2022).

[47] W. Zhang, H. Yuan, N. Sun, H. Sun, and X. Zhang, Nat. Commun. **13**, 2937 (2022).

[48]　See Supplemental Materials for the derivation of circuit eigen-equations, the details for sample fabrications and circuit signal measurements, as well as the time-domain simulations.

**Acknowledgements**. This work was supported by the National Key R & D Program of China under Grant No. 2017YFA0303800 and the National Natural Science Foundation of China (No.12104041).

**Competing interests.** The authors declare no competing interests.